\documentclass[prb,aps,twocolumn,showpacs,amsmath,amssymb]{revtex4}

\usepackage{graphicx}
\usepackage{dcolumn}
\usepackage{bm}

\begin{document}
\title{Giant Anisotropy of  Magnetoresistance and "Spin Valve" effect in Antiferromagnetic $Nd_{2-x}Ce_xCuO_{4}$}
\author{T. Wu$^1$, C. H. Wang$^1$, G. Wu$^1$, D. F. Fang$^1$, J. L. Luo$^2$, G T. Liu$^2$ and X. H. Chen$^{1,\ast}$}
\affiliation{1. Hefei National Laboratory for Physical Science at
Microscale and Department of Physics, University of Science and
Technology of China, Hefei, Anhui 230026, People's Republic of
China\\ 2. Beijing National Laboratory for Condensed Matter Physics,
Institute of Physics, Chinese Academy of Science, Beijing 100080,
People's Republic of China}

\date{\today}

\begin{abstract}
We have studied anisotropic magnetoresistance (MR) and
magnetization with rotating magnetic field (B) within $CuO_2$
plane in lightly doped AF $Nd_{2-x}Ce_xCuO_{4}$. \emph{A giant
anisotropy} in MR is observed at low temperature below 5 K. The
c-axis resistivity can be tuned about one order of magnitude just
by changing B direction within $CuO_2$ plane and a scaling
behavior between out-of-plane and in-plane MR is found. A "Spin
valve" effect is proposed to understand the giant anisotropy of
out-of-plane MR and the evolution of scaling parameters with the
external field. It is found that the field-induced spin-flop
transition of Nd$^{3+}$ layer under high magnetic field is the key
to understand the giant anisotropy. These results suggest that a
novel entanglement between charge and spin dominates the
underlying physics.
\end{abstract}


\maketitle
\newpage
\section{Introduction}
 It is generally believed that the pairing necessary for
high-$T_c$ superconductivity in cuprates involves the interplay
between doped charges and antiferromagnetic (AF) spin correlation.
In this sense, the study of lightly doped, insulating AF state is
important to understand the pairing mechanism because density of
the carriers can be sufficiently low so that the interaction
between them is small relative to their interaction with $Cu^{2+}$
spins. Many intriguing and anomalous phenomena were observed in
lightly doped AF cuprates due to strong coupling between charges
and $Cu^{2+}$ spins.\cite{thio,ando1,ando2,ando3,chen} $Cu^{2+}$
spins order in an AF collinear structure for the parent compounds
of hole-doped cuprates,\cite{vaknin,tranquada1} while in AF
noncollinear structure for that of electron-doped
cuprates.\cite{skanthakumar,sumarlin} All spins point either
parallel or antiparallel to a single direction in AF collinear
structure, while the spins in adjacent layers are orthogonal in AF
noncollinear structure. A transition from noncollinear to a
collinear spin arrangement with a spin-flop can be induced by
certain magnetic field ($B_c$)\cite{plakhty}, which is confirmed
in lightly electron-doped
$Pr_{1.3-x}La_{0.7}Ce_xCuO_4$\cite{ando3} and
$Nd_{2-x}Ce_xCuO_4$\cite{chen} crystals, and such transition
affects significantly both the in-plane and out-of-plane
resistivity.

In $Nd_2CuO_4$, the $Cu^{2+}$ spins order in three phases with two
different AF noncollinear spin structures and experience two
reorientation phase transitions
\cite{skanthakumar1,matsuda,skanthakumar,skanthakumar2}. It has
been reported by us that MR anisotropy with a fourfold symmetry in
different AF spin structures upon rotating magnetic field (B)
within ab-plane, while with a twofold symmetry at the spin
reorientation temperatures, is observed in lightly doped
$Nd_{2-x}Ce_xCuO_4$ above 10 K.\cite{chen} A large anisotropic MR
was observed in lightly electron-doped
$Pr_{1.3-x}La_{0.7}Ce_xCuO_4$\cite{ando3} and
$Nd_{2-x}Ce_xCuO_4$\cite{chen}. These results indicate strong
spin-charge coupling in electron-doped cuprates. In Nd$_2$CuO$_4$,
the magnetic coupling between Nd$^{3+}$ and Cu$^{2+}$ is very
important at low temperature since the magnetic moment of
Nd$^{3+}$ becomes large with decreasing temperature ($\sim$
1.3$\mu_B$ at 0.4 K).\cite{matsuda} Magnetic structure of
Nd$^{3+}$ is very abundant at low temperature.\cite{Cherny,
Richard1, Richard2} In this sense, electronic transport at low
temperature is expected to be sensitive for the change of magnetic
structure of Nd$^{3+}$ due to strong spin-charge coupling. It will
provide us a chance to understand the spin-charge coupling in
electron-doped cuprates. The lightly electron-doped cuprates are
good system to study the coupling between charge and $Cu^{2+}$
spin because: (1) the spin structure can be tuned by external
magnetic field;\cite{plakhty} (2) in contrast to the buckling of
$CuO_2$ in hole doped cuprates, the $CuO_2$ plane in
electron-doped cuprates is flat, so that the spin ordering is pure
antiferromagnetic without ferromagnetic component along c-axis
occurred in hole doped cuprates, such ferromagnetic component
along c-axis makes the study of the coupling between charge and
$Cu^{2+}$ spin complicated. In this work, we study angular
dependent magnetoresistance and magnetization below 10 K in
lightly electron-doped $Nd_{2-x}Ce_xCuO_4$. \emph{A giant
anisotropy} in MR is observed, and the c-axis resistivity can be
tuned about one order of magnitude just by changing B direction.
Scaling behavior between in-plane and out-of-plane MR is
systematically changed with increasing magnetic field. The jump in
MR with B around Cu-O-Cu direction coincides with the sudden
change in magnetization at low temperature below 5 K. The
underlying physics will be discussed below.

\section{Experiment}
Growth of single crystals and their resistivity have been reported
in previous work.\cite{chen} Susceptibility and magnetoresistance
were measured with the superconducting quantum interference device
(SQUID) with 7 Tesla maximal magnetic field and quantum design
PPMS system with 12 Tesla maximal magnetic field, respectively. In
our measurements, the maximal magnetic field is 12 Tesla for MR
and 7 Tesla for magnetization. The $\rho$$_{ab}$ and $\rho$$_c$
stand for in-plane resistivity and out-of-plane resistivity,
respectively. The magnetoresistance is defined as
MR=$\frac{\triangle\rho(B)}{\rho(0)}$=$\frac{\rho(B)-\rho(0)}{\rho(0)}$.
It should be addressed that all results discussed as follow are
well reproducible.

\section{Result and Discussion}
Fig.1 shows the isothermal out-of-plane MR at 5 K for the single
crystals with x=0.025 and 0.033 with B along Cu-Cu and Cu-O-Cu
direction, respectively. The MR behavior is similar to that
observed in antiferromagnetic $Pr_{1.3-x}La_{0.7}Ce_xCuO_4$ with
x=0.01 crystal.\cite{ando3} But the magnitude of MR and the MR
anisotropy are much larger than the case of
$Pr_{1.3-x}La_{0.7}Ce_xCuO_4$. The step-like increase of MR
corresponds to the noncollinear-collinear transition occurring at
the critical field $B_c$. As shown in Fig.1, the critical field
$B_c$ along Cu-O-Cu direction is larger than that along Cu-Cu
direction. Above $B_c$, the behavior of MR for B along Cu-Cu
direction is quite different from that for B along Cu-O-Cu
direction in the collinear structure. The MR with B along Cu-Cu
direction slightly changes above $B_c$, while the MR monotonically
increases with increasing B for B along Cu-O-Cu direction. A giant
anisotropic MR between B along Cu-Cu and Cu-O-Cu direction is
observed. For x=0.025 crystal, the MR at 12 T is as high as $\sim
235 \%$  with B along Cu-O-Cu direction, while only $\sim 17 \%$
with B along Cu-Cu direction.

\begin{figure}[t]
\includegraphics[width=9cm]{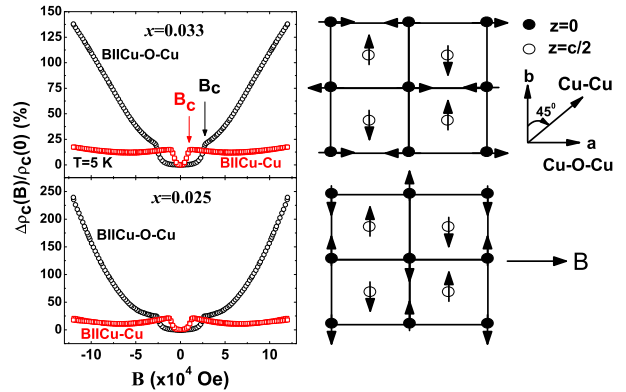}
\caption{Isothermal MR at 5 K with the B along Cu-O-Cu and Cu-Cu
direction for the samples $Nd_{2-x}Ce_xCuO_4$ with x=0.025 and
0.033, respectively. Zero-field noncollinear spin structure, only
Cu spins are shown; Field-induced transition from noncollinear to
collinear spin ordering with B along Cu-O-Cu direction.\\}
\end{figure}

Upon rotating a magnetic field larger than $B_c$ within the
$CuO_2$ plane, the spins always keep the collinear arrangement and
the spin structure rotates as a whole, all spins are perpendicular
to the magnetic field as shown in Fig.1.\cite{plakhty} In order to
study the anomalous and giant anisotropic MR, we carefully
investigated the evolution of in-plane and out-of-plane MR with
rotating B within $CuO_2$ plane.  Fig. 2a and 2b show the
evolution of in-plane and out-of-plane MR with the angle between B
and Cu-O-Cu ([100]) direction at 5 K for the single crystal with
x=0.025. Both of in-plane and out-of-plane MR increase with
increasing B, and show a giant anisotropy with fourfold-symmetry,
such fourfold symmetry arises from the symmetry of magnetic
structure because there exist two equivalent spin easy axes (Cu-Cu
direction) and two equivalent spin hard axes (Cu-O-Cu direction)
in the collinear spin structure, which has been confirmed by the
different critical fields $B_c$ for B along Cu-O-Cu and Cu-Cu
directions as shown in Fig.1. A striking feature is observed that
the out-of-plane MR at 12 T sharply increases from $\sim 200\%$ to
$\sim 300 \%$ at the angle close to B along Cu-O-Cu. Such behavior
originates from the spin-flop induced by magnetic field with B
close to the Cu-O-Cu direction as discussed below. A similar jump
can be also observed at the angle close to B along Cu-Cu, but the
jump is very small compared to the case of B close to Cu-O-Cu
direction.

\begin{figure}[t]
\includegraphics[width=0.4\textwidth]{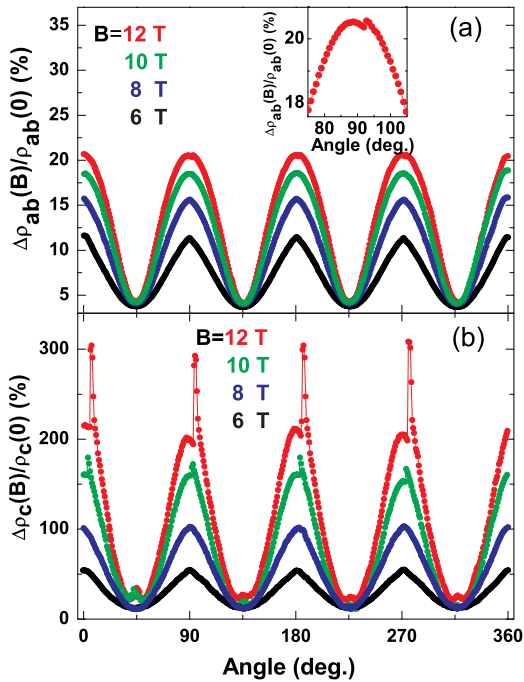}
\caption{(a): Isothermal in-plane and (b): Out-of-plane MR at 5 K
under different B as a function of angle between B and Cu-O-Cu
direction upon rotating B within $CuO_2$ plane for the single
crystal with x=0.025. The inset in (a): magnified in-plane MR with
B = 12 T.\\}
\end{figure}

\begin{figure}[t]
\includegraphics[width=0.4\textwidth]{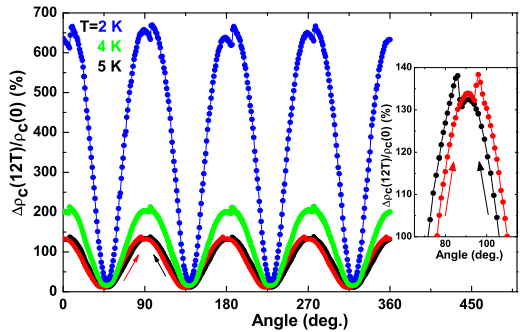}
\caption{Out-of-plane MR as a function of angle between B and
Cu-O-Cu direction upon rotating B within $CuO_2$ plane at 2K, 4K and
5 K, respectively, for the single crystal with x=0.033 (B=12 T).\\}
\end{figure}

In order to study the effect of temperature on the anisotropy of
MR, we systematically investigated the MR behavior of the x=0.033
crystal because resistivity of the x=0.025 crystal is too large to
be measured due to the resistivity divergence at low temperature.
Fig.3 shows evolution of the out-of-plane MR upon rotating B
within $CuO_2$ plane at 2, 4 and 5 K under 12 T for the x=0.033
crystal. The results are similar to that observed in the crystal
with x=0.025. The MR increases monotonically and the anisotropy of
MR induced by rotating B within Cu-O plane apparently increases
with decreasing temperature. The MR under 12 T with B along Cu-Cu
direction is about 11.2\% at 5 K, 17.1\% at 4 K and 27.7\% at 2 K;
while the MR with B along Cu-O-Cu direction is about 133\% at 5 K,
203\% at 4 K and 656\% at 2 K, respectively. It indicates that a
giant anisotropy of resistivity is induced by magnetic field with
B along Cu-O-Cu and Cu-Cu at low temperature. At 2 K, the
resistivity under 12 T with B along Cu-O-Cu direction is about one
order of magnitude larger than that with B along Cu-Cu direction.
Such giant anisotropy in resistivity induced just by changing B
direction within $CuO_2$ plane should be related to the magnetic
structure and magnetic moment induced by B because the magnetic
field along Cu-O-Cu or Cu-Cu just changes the spin structure and
induces the different magnitude of the magnetic moment. To
understand the jump in MR with B close to the Cu-O-Cu direction,
the MR at 5 K is measured with rotating B in clockwise direction
and in anti-clockwise direction, respectively. It is found that
the MR jumps observed with rotating B in clockwise direction and
in anti-clockwise direction are symmetric relative to the B along
Cu-O-Cu direction as shown in the inset of Fig.3. It indicates
that the spin does not prefer to the Cu-O-Cu direction, and the
spin jump always occurs around Cu-O-Cu direction when B is rotated
within $CuO_2$ plane. Therefore, the jump arises from the
spin-flop induced by B.

\begin{figure}[t]
\includegraphics[width=0.4\textwidth]{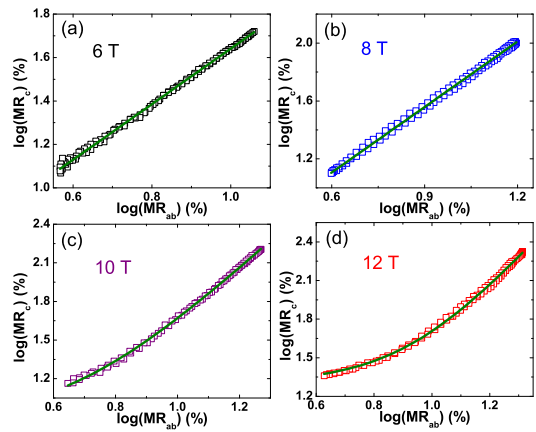}
\caption{The same data shown in Fig.2 are plotted in $\Delta
\rho_{c}(B)/(\rho_{c}(0)$) as a function of $\Delta
\rho_{ab}(B)/(\rho_{ab}(0)$). The line is the fitting results with
the formula $\Delta\rho_{c}(B)/\rho_{c}(0)$=
$\beta+\alpha(\Delta\rho_{ab}(B)/\rho_{ab}(0)$)$^\upsilon$. At 6 T
and 8 T, the parameter $\beta$ = 0.\\}
\end{figure}

\begin{table}[h]
\caption{\label{tab:table1} Fitting parameters $\alpha$, $\beta$ and
$\upsilon$ with the formula $\Delta\rho_{c}(B)/\rho_{c}(0)$=
$\beta+\alpha(\Delta\rho_{ab}(B)/\rho_{ab}(0)$)$^\upsilon$ under
different field.}
\begin{ruledtabular}
\begin{tabular}{cccc}
Field & $\beta$ & $\alpha$ & $\upsilon$\\
\hline 6 T & 0 & 2.30 & 1.29 \\
8 T & 0 & 1.59 & 1.51 \\
10 T & 7.47 & 0.25 & 2.20 \\
12 T & 20.13 & 0.10 & 2.50 \\
\end{tabular}
\end{ruledtabular}
\end{table}

As shown in Fig.4, the same data of in-plane and out-of-plane MR
shown in Fig.2a and 2b are plotted in
$\Delta\rho_{c}(B)/\rho_{c}(0)$ as a function of
$\Delta\rho_{ab}(B)/\rho_{ab}(0)$. Only the data of in-plane and
out-of-plane MR between 45 and 90 degree is plotted in Fig.4
because in-plane and out-of-plane MR exhibit the exactly same
oscillation. All data above can be fitted by
MR$_c$=$\beta$+$\alpha$$\cdot$MR$^\upsilon_{ab}$ very well.  The
fitting parameters are list in Table I. It is found that the
fitting parameter $\beta$ is zero below 10 T. The fitting
parameter $\upsilon$ increases from $\sim$1 to $\sim$3 with
increasing magnetic field. These results indicate that the
relation between out-of-plane MR and in-plane MR is strongly
dependent on the external magnetic field. It is suggested that the
out-of-plane and in-plane transport is closely related to magnetic
structure since the external magnetic field can modify the spin
structure. The giant anisotropy could arise from the change of
spin structure induced by external magnetic field.

\begin{figure}[b]
\includegraphics[width=9cm]{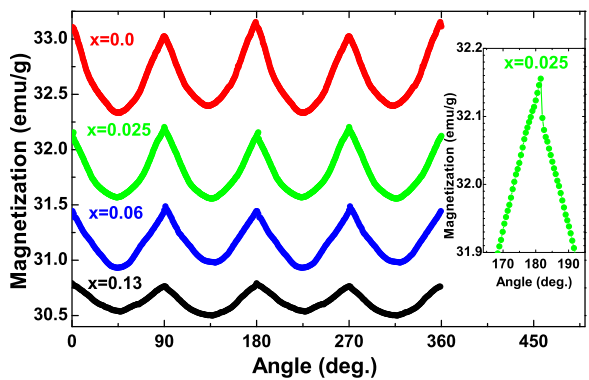}
\caption{The magnetization measured under magnetic field of 7 Tesla
at 2 K as a function of angle between B and Cu-O-Cu direction upon
rotating B within $CuO_2$ plane for the samples with x=0, 0.025,
0.06, 0.13. Inset shows a jump in magnetization at certain angle
corresponding to the jump observed in MR.\\}
\end{figure}

\begin{figure}[h]
\includegraphics[width=0.4\textwidth]{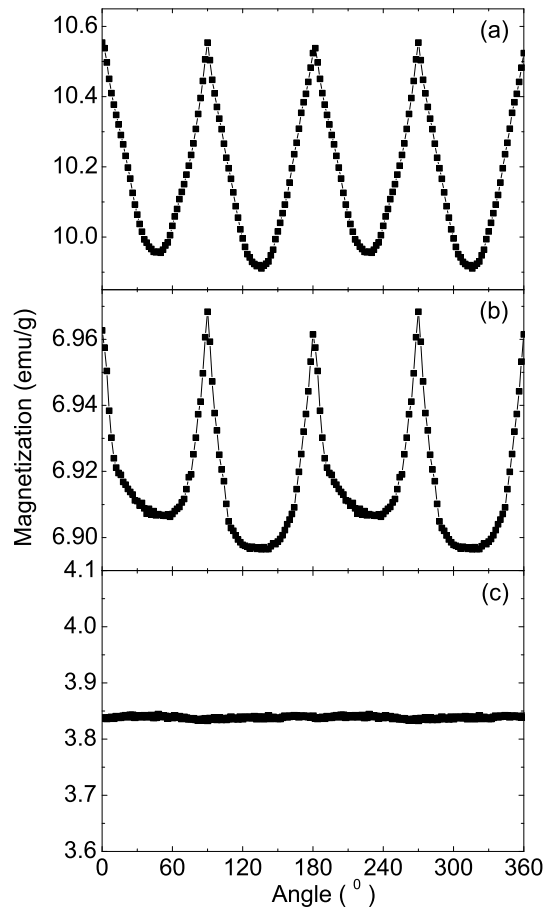}
\caption{The increment of magnetization measured under magnetic
field of 6.5 Tesla at (a): 2 K, (b): 3.5 K and (c): 5 K relative to
the magnetization at 7.5 K as a function of angle between B and
Cu-O-Cu direction upon rotating B within $CuO_2$ plane for the
samples with x= 0.025.\\}
\end{figure}

\begin{figure}[h]
\includegraphics[width=0.5\textwidth]{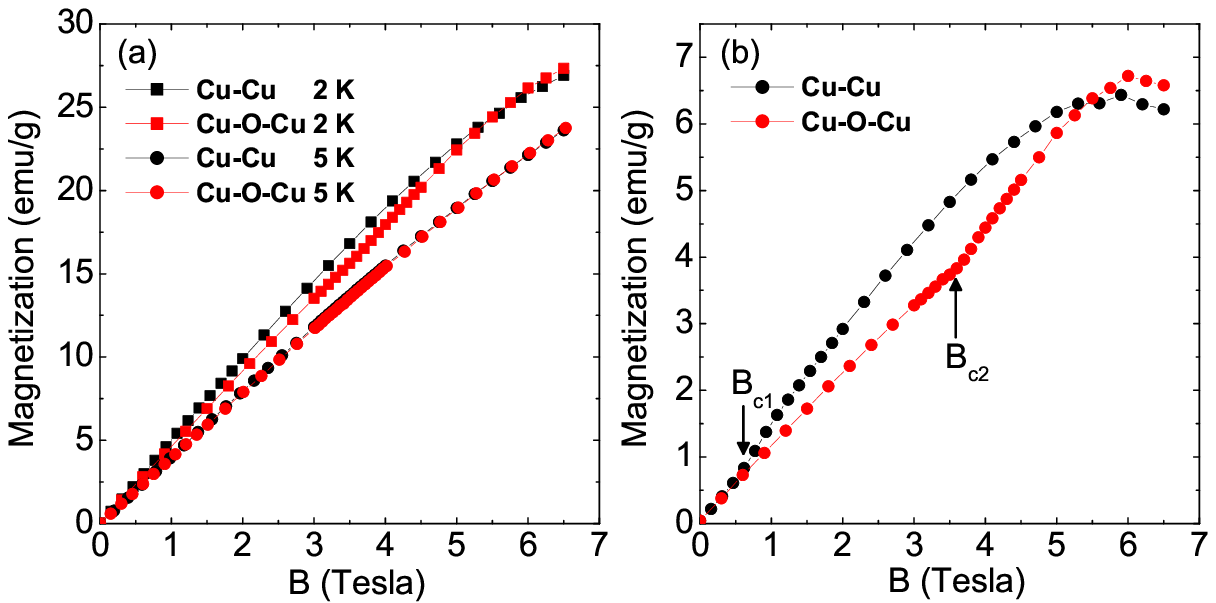}
\caption{(Color online) (a): Magnetization as a function of magnetic
field along Cu-Cu and Cu-O-Cu direction at 2 K and 5 K; (b):
M(2K)-M(5K).}
\end{figure}

In order to further understand how magnetic field influences the
transport, understanding on the evolution of magnetic structure
under magnetic field is very necessary. Fig.5 shows the
magnetization under 7 Tesla at 2 K with rotating B within $CuO_2$
plane for the AF $Nd_{2-x}Ce_xCuO_2$ with x=0, 0.025, 0.06 and
0.13. It is found that magnetization shows the same fourfold
symmetry with rotating B within $CuO_2$ plane as that observed in
MR shown in Fig.2. The amplitude of the oscillation and the
magnetization decrease with increasing x. A striking feature is
observed that the magnetization shows a jump with B around Cu-O-Cu
direction at which a corresponding jump is observed in MR as shown
in the inset of Fig.5. However, this fourfold symmetry gradually
disappears with increasing temperature as shown in Fig.6. As we
know, the magnetic moment of Nd$^{3+}$ increases prominently at
low temperature, and the magnetization is very sensitive to
magnetic moment of Nd$^{3+}$ below 5 K\cite{Cherny, Lynn}.
Therefore, the fourfold symmetry in magnetization and the jump in
magnetization are related to the magnetic structure of Nd$^{3+}$.
In the other hand, the fourfold symmetry shown in Fig.5 indicates
a magnetic ordering of Nd$^{3+}$. The spontaneous ordering of the
Nd$^{3+}$ subsystem at low temperature due to Nd$^{3+}$-Nd$^{3+}$
interaction remains controversial. X-ray magnetic scattering data
indicate that Nd$^{3+}$ ions are polarized at 37 K\cite{Hill}. The
removal of the Kramers doublet degeneracies observed by crystal
field infrared transmission indicates that these ions are already
polarized by the Cu$^{2+}$ subsystem at a temperature as high as
140 K\cite{Jandl}. An enhancement of neutron scattering magnetic
peak intensities around 3 K has been interpreted as Nd$^{3+}$
ordering due to Nd$^{3+}$-Nd$^{3+}$ interaction\cite{matsuda},
while Lynn et al. have estimated the Nd$^{3+}$ ordering
temperature around 1.5 K\cite{Lynn}. Recently, an abnormal peak
around 5 K observed in ultrasonic measurement is explained to be
somehow related to local magnetic domains\cite{Richard1}. Since
the Nd$^{3+}$-Cu$^{2+}$ and Nd$^{3+}$-Nd$^{3+}$ interactions are
opposite, the former is dominated above 5 K and makes Nd$^{3+}$
parallel to Cu$^{2+}$ as shown in Fig.8(a), while the later is
dominated below 5 K and makes Nd$^{3+}$ prefer to be perpendicular
to Cu$^{2+}$ as shown in Fig.8(b). Due to the frustration of the
Nd$^{3+}$ magnetic subsystem arising from the competition between
Nd$^{3+}$-Cu$^{2+}$ and Nd$^{3+}$-Nd$^{3+}$ interaction, the local
magnetic domain is formed with Nd$^{3+}$ not parallel to Cu$^{2+}$
magnetic moment below 5 K. The magnetic structure of Nd$^{3+}$
subsystem with magnetic moment of Nd$^{3+}$ perpendicular to
Cu$^{2+}$ can be stabilized by the external field. The observed
fourfold symmetry below 5K in magnetization could arise from the
reorientation of Nd$^{3+}$ spin. Richard et al. have given
evidence that the magnetic structure below 5 K has anisotropic
field-dependence\cite{Richard1}. As shown in Fig.7, the field
dependent magnetization of x=0.025 sample at 2 K shows an anomaly
at 0.6 T and 3.6 T for both Cu-Cu and Cu-O-Cu directions,
respectively. However, no such anisotropy is observed above 5 K.
To make the anisotropy clear, the magnetization at 2 K subtracted
the 5 K magnetization is shown in Fig.7(b). Such anomaly has been
attributed to spin-reorientation of Nd$^{3+}$ in
Nd$_2$CuO$_4$\cite{Cherny}. The spin reorientation of Nd$^{3+}$
occurs due to a transition from the magnetic structure shown in
Fig.8(a) to that shown in Fig.8(b). It is surprising that the
magnetization for two directions has a cross around 6 T and the
magnetization shows somehow saturation as shown in Fig. 7. It is
suggested that the magnetic structure under high magnetic field is
different from that under low magnetic field. Similar result has
been reported in Nd$_{2}$CuO$_4$\cite{Cherny}. Recently, a
crossover from antiferromagnetic to paramagnetic configuration
induced by high magnetic field is proposed by Richard et
al.\cite{Richard2}. The corresponding magnetic structures given by
Richard et al. are shown in Fig.8(c)-(f). When in-plane magnetic
field B $<$ 4 T along Cu-O-Cu and Cu-Cu directions, the collinear
magnetic structures are shown in Fig.8(c) and (e), respectively.
In the configuration, the Cu$^{2+}$-Nd$^{3+}$ interaction is
larger than Nd$^{3+}$-Nd$^{3+}$ interaction. When in-plane
magnetic field B $>$ 4 T along Cu-O-Cu and Cu-Cu directions, the
Nd$^{3+}$-Nd$^{3+}$ interaction is dominated and the magnetic
structure changes from antiferromagntic to paramagnetic
configuration as shown in Fig.8 (d) and (f), in which the
Nd$^{3+}$ spins are aligned in applied magnetic field and thus
behave as ferromagnetic-like. The cross at about 6 T in
magnetization could be related to the change of magnetic structure
shown in Fig.8.

\begin{figure}[h]
\includegraphics[width=0.4\textwidth]{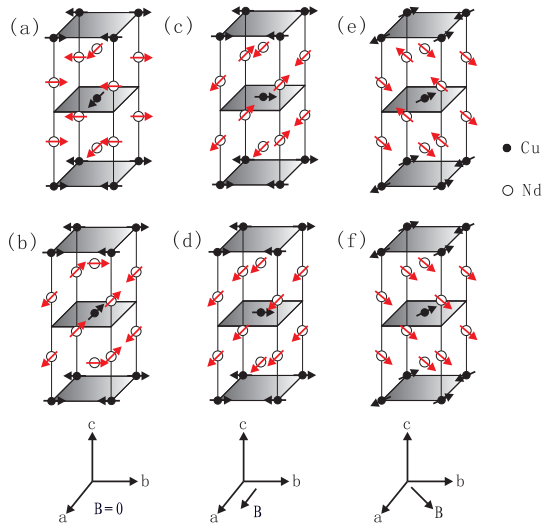}
\caption{(Color online) Magnetic configuration of Nd$_2$CuO$_4$ in
the noncollinear antiferromagnetic phase (a): above 5 K; (b):
below 5 K; Magnetic configuration of Nd$_2$CuO$_4$ in the
collinear phase assuming an antiferromagnetic alignment of
Nd$^{3+}$ spins (c): B$\parallel$[100]; (e): B$\parallel$[110]],
and a ferromagnetic-like alignment of the Nd$^{3+}$ spins (d):
B$\parallel$[100]; (f): B$\parallel$[110].}
\end{figure}

The fourfold symmetry in magnetoresistance has been observed above
5 K in the same sample in previous result\cite{chen}, while
similar symmetry in magnetization arose from Nd$^{3+}$ spin is
observed only below 5 K. Therefore, the fourfold symmetry in MR
should arise from anisotropic magnetic structure of $Cu^{2+}$ as
discussed in our previous work\cite{chen}. As shown in Fig.2 and
Fig.6, the giant anisotropy in MR below 5 K coincides with the
magnetic ordering of Nd$^{3+}$ with the same fourfold symmetry.
These results indicate that the fourfold symmetry of MR results
from  spin ordering of $Cu^{2+}$, and the spin ordering of
Nd$^{3+}$ enhances the fourfold symmetry in MR below 5 K and leads
to a giant anisotropy in MR. It is evident that the jump of
out-of-plane MR with B along Cu-O-Cu direction shown in Fig.2 is
related to the sudden change in magnetization shown in the inset
of Fig.5.  Therefore, the change of MR below 5 K relative to high
temperature MR can be mainly ascribed to the ordering of Nd$^{3+}$
spin. As shown in Fig.4, there exists a scaling behavior between
the out-of-plane MR and in-plane MR with
MR$_c$=$\beta$+$\alpha$$\cdot$MR$^\upsilon_{ab}$.  But the fitting
parameters $\beta$, $\alpha$ and $\upsilon$ show a systematic
change with increasing external field as listed in Table 1. Change
of the scaling behavior is closely related to the change of
magnetic structure of Nd$^{3+}$ induced by external field since
the increase of external field from 6 T to 12 T cannot lead to
change of magnetic structure of $Cu^{2+}$ subsystem. It is
supported by the fact that the change of in-plane MR with
increasing magnetic field is much smaller than that of
out-of-plane MR, and the anisotropy of out-of-plane MR is much
larger than that of in-plane MR as shown in Fig.2. Richard et
al.,\cite{Richard2} pointed out that the magnetic structure of
Nd$^{3+}$ subsystem can change from antiferromagnetic to
paramagnetic configuration at certain critical magnetic field. It
is possible that the evolution of scaling parameters listed in
Table 1 is related to the change of magnetic structure of
Nd$^{3+}$. It is well known that the neighboring $CuO_2$ plane
with antiferromagnetic configuration is separated by Nd-O
layer.\cite{Richard1} As discussed above, the magnetic structure
of Nd$^{3+}$ subsystem can be turned by external field. Therefore,
the out-of-plane transport can be switched by magnetic field
assuming the Nd-O layer as a barrier. In this sense, this
phenomenon can be well understood with "Spin Valve" effect. The
magnetic excitations are different with B along Cu-Cu and Cu-O-Cu
directions because the magnetic structure is more frustrated
around Cu-O-Cu\cite{Richard1}. In "Spin Valve" picture, the
different magnetic excitations in Nd$^{3+}$ layer lead to
different transport along c-axis. Therefore, the spin-flop
transition is the key to understand the giant anisotropy of
out-of-plane MR. Thermal conductivity results indicate that
in-plane magnetic field can result in a close of anisotropic gap
($\sim$ 0.3meV) which leads to additional heat
conduction\cite{Jin}. The critical magnetic field is about 4.5 T
and 2.5 T in Cu-O-Cu and Cu-Cu direction, respectively. The
closure of anisotropic gap corresponds to spin-flop transition for
Cu$^{2+}$ spin. This result indicates that the spin-flop
transition can lead to a closure of anisotropic gap. At low
temperature below 5 K, another gap related to spin-flop transition
of Nd$^{3+}$ is closed under high magnetic field and this gap is
anisotropic.  The gap can be estimated with B=$\Delta$/g$\mu$$_B$
($B\sim8 T$), the gap is about 0.5 meV with magnetic field along
Cu-O-Cu direction. This spin-flop should originate from Nd$^{3+}$
subsystem because magnons from $Cu^{2+}$ have energy above 5
meV\cite{Bourges, Petitgrand, Ivanov}, and four optical Nd magnon
branches lie in the range 0.2 to 0.8 meV\cite{Henggeler, Casalta}.
The study on magnetic structure under high magnetic field at low
temperature is lacking. This picture needs further experimental
investigation to confirm. These results give a strong evidence
that a nontrivial correlation between charge and AF ordering
background exists, the charge transport can be affected not only
by Cu$^{2+}$ spins but also Nd$^{3+}$ spins, especially below 5 K.

The huge changes in resistivities as induced by the in-plane
magnetic field seem to be a highly nontrivial phenomenon. Note
that the applied in-plane magnetic field should only affect the
spins of the system via the Zeeman effect without directly
influencing the orbital motion of charge carriers in the in-plane
case, and presumably with only a weak orbital effect for the
out-of-plane case as the resistivity itself is divergent at low
temperature. It thus implies the existence of some kind of strong
``entanglement" between the spin and charge degrees of freedom
such that by tuning the magnetic ordering with an in-plane
magnetic field can result in a big enhancement of resistivities
seen in the measurements. Furthermore, the large MR behavior in
this insulating regime also strongly suggests that the divergence
of resistivities at low-temperature may not be simply a
conventional localization effect due to disorders since spin
structures can affect resisitivities so much. Although the
microscopic mechanism remains unclear, the novel spin-charge
entanglement does exist in strongly correlated models. For
example, in the t-J model, a so-called phase string effect has
been shown \cite{phase string} to be present as a non-local mutual
frustration between the charge and spin degrees of freedom induced
by doped charge carriers moving in an antiferromanget. In fact,
the localization of the charge carriers in the magnetic ordered
phase has been interpreted\cite{localization} based on such a
phase string effect and it is thus conceivable that the change of
the spin structure may strongly affect the resistivities via the
phase string effect. The scaling between in-plane and out-of-plane
MR is strongly dependent on spin structure of Nd$^{3+}$ which
emerges at low temperature since strong Cu$^{2+}$-Nd$^{3+}$
interaction and can be easily tuned by external magnetic field. It
provides a good chance to show a evidence for the spin-charge
entanglement.

\section{Conclusion}
In this paper, we study anisotropic magnetoresistance (MR) and
magnetization with rotating magnetic field (B) within $CuO_2$ plane
in lightly doped AF $Nd_{2-x}Ce_xCuO_{4}$. \emph{A giant anisotropy}
in MR is observed, and the c-axis resistivity can be tuned by about
one order of magnitude just with changing B direction. These results
provide evidence to support the spin-flop transition of Nd$^{3+}$
ions induced by high magnetic field. The change of magnetic
structure induced by different external field leads to a systematic
evolution of the scaling behavior between in-plane and out-of-plane
MR. "Spin Valve" effect is used to well understand the out-of-plane
MR behavior. Such novel entanglement of charge and spin dominates
the underlying physics.

\vspace*{2mm} {\bf Acknowledgment:} The authors would like to
thank Z. Y. Weng for stimulating discussions and constructive
opinion. We acknowledge T. Xiang and Q. H. Wang for helpful
discussions. This work is supported by the Nature Science
Foundation of China and by the Ministry of Science and Technology
of China (973 project No: 2006CB601001) and by National Basic
Research Program of China (2006CB922005).

$^{\ast}$ Corresponding author, \emph{Electronic address:}
chenxh@ustc.edu.cn

\end{document}